\documentclass[prd,twocolumn,showpacs,nofootinbib]{revtex4}

\usepackage{amsmath}
\usepackage{graphicx}
\usepackage{dcolumn}
\usepackage{bm}
\usepackage{amssymb}
\usepackage{latexsym}
\usepackage{subfigure}
\newcommand{\ltsim}{\mathrel{\raise.3ex\hbox{$<$\kern-.75em\lower1ex\hbox{$\sim$}}}}
\newcommand{\gtsim}{\mathrel{\raise.3ex\hbox{$>$\kern-.75em\lower1ex\hbox{$\sim$}}}}
\newcommand{\be}{\begin{equation}}
\newcommand{\ee}{\end{equation}}
\newcommand{\bea}{\begin{eqnarray}}
\newcommand{\eea}{\end{eqnarray}}

\def\mpch{\,{h {\rm Mpc}^{-1}}}

\def\der{{\rm d}}

\begin{document}

\title{How much cosmological information can be measured?}

\author{Yin-Zhe Ma$^{1,2}$\footnote{Email: ma@ukzn.ac.za} and Douglas Scott$^2$\footnote{Email: dscott@phas.ubc.ca}}

\affiliation{$^{1}$School of Chemistry and Physics, University of KwaZulu-Natal, Westville Campus, Private Bag X54001, Durban, 4000, South Africa
\\
$^{2}$Department of Physics and Astronomy, University of British Columbia,
 6224 Agricultural Road, Vancouver, British Columbia, Canada. V6T 1Z1}

% ----------------------- ABSTRACT -------------------------

\begin{abstract}
It has become common to call this the ``era of precision cosmology,''
and hence one rarely hears about the finiteness of the amount of
information that is available for constraining cosmological
parameters.  Under the assumption that the perturbations are
purely Gaussian, the amount of extractable information (in terms
of total signal-to-noise ratio for power spectrum measurements) is the
same (up to a small numerical factor) as an accounting of the
number of observable modes.  For studies of the microwave sky, we
are probably within a factor of a few of the amount of accessible
information.  To dramatically reduce the uncertainties on
parameters will require three-dimensional probes, such as ambitious future
redshifted 21-cm surveys.  However, even there the
available information is still finite, with the total effective
signal-to-noise ratio on parameters probably not exceeding $10^7$.
The amount of observable information will increase with time (but very
slowly) into the extremely distant future.
\end{abstract}


\maketitle

% ---------------------- INTRO -------------------------------------------

\section{Introduction}
The standard cosmological model has been
confirmed with ever-increasing precision using cosmic microwave background
(CMB) data, such as from the {\it Wilkinson Microwave Anisotropy
Probe}~\cite{Hinshaw12} and now particularly from the
{\it Planck\/}
satellite~\cite{Planck2013I,PlanckI}. This model (hereafter referred
to as $\Lambda$CDM), containing a cosmological constant
($\Lambda$) and cold dark matter, is built on a framework of
simplifying assumptions within which only a relatively small
number of free parameters are required.  One can
consider that the information contained in the data is compressed
down to the combined constraints on these cosmological parameters.
To the extent that the observable Universe comes from a Gaussian random
process, all of the information is contained in the two-point functions (which
are limited by cosmic variance), and not in the particular realization of the
CMB sky (which can in principle be determined to arbitrary precision).

As the constraining power of the data improves, the parameters are
determined more precisely, and hence for the parameter set some overall
``signal-to-noise'' increases with time.  Currently we have tight constraints
on a six-parameter $\Lambda$CDM model, with the parameters conventionally chosen
to be the baryon density, $\Omega_{\rm b}h^2$, with $h$ being the Hubble
constant in units of $100\,{\rm km}\,{\rm s}^{-1}\,{\rm Mpc}^{-1}$; the
cold dark matter density, $\Omega_{\rm c}h^2$; the angular size of
the sound horizon, $\theta_\ast$, which is a function of the
$\Omega$s; the amplitude of the power spectrum of initial density
perturbations, $A_{\rm s}$; the slope of this power spectrum,
$n_{\rm s}$; and the optical depth due to reionization, $\tau_{\rm ion}$.
Improvements in data quality are leading to tighter and tighter
constraints on these parameters, and hence one is led to consider
whether there is some ultimate precision that can be obtained, or
alternatively one can ask how much cosmological information can be
measured.

In fact, {\it Planck\/} has already mapped a large fraction of the
primordial anisotropies in temperature~\cite{Planck16} and hence we are
starting to reach the point where we are squeezing the CMB sky for the
remains of the information (although this is {\it not\/} yet true for
polarization, see Ref.~\cite{2dinfo}).
To obtain a dramatic increase in cosmological information
therefore requires us to consider the observable
modes of the three-dimensional power spectrum, $P_{\rm m}(k)$, as measured using
galaxy clustering, cosmic shear, or 21-cm correlations, for example.

\section{Size of the observable Universe and the far future}
We first investigate the maximal space that can be observed and the
smallest linear-scale structure one can probe at different epochs.
Then we calculate the total number of
(linear) modes as a function of cosmic time, and the ultimate
parameter precision one can achieve.

\begin{figure*}[tbp!]
\centerline{
\includegraphics[width=4.5in]{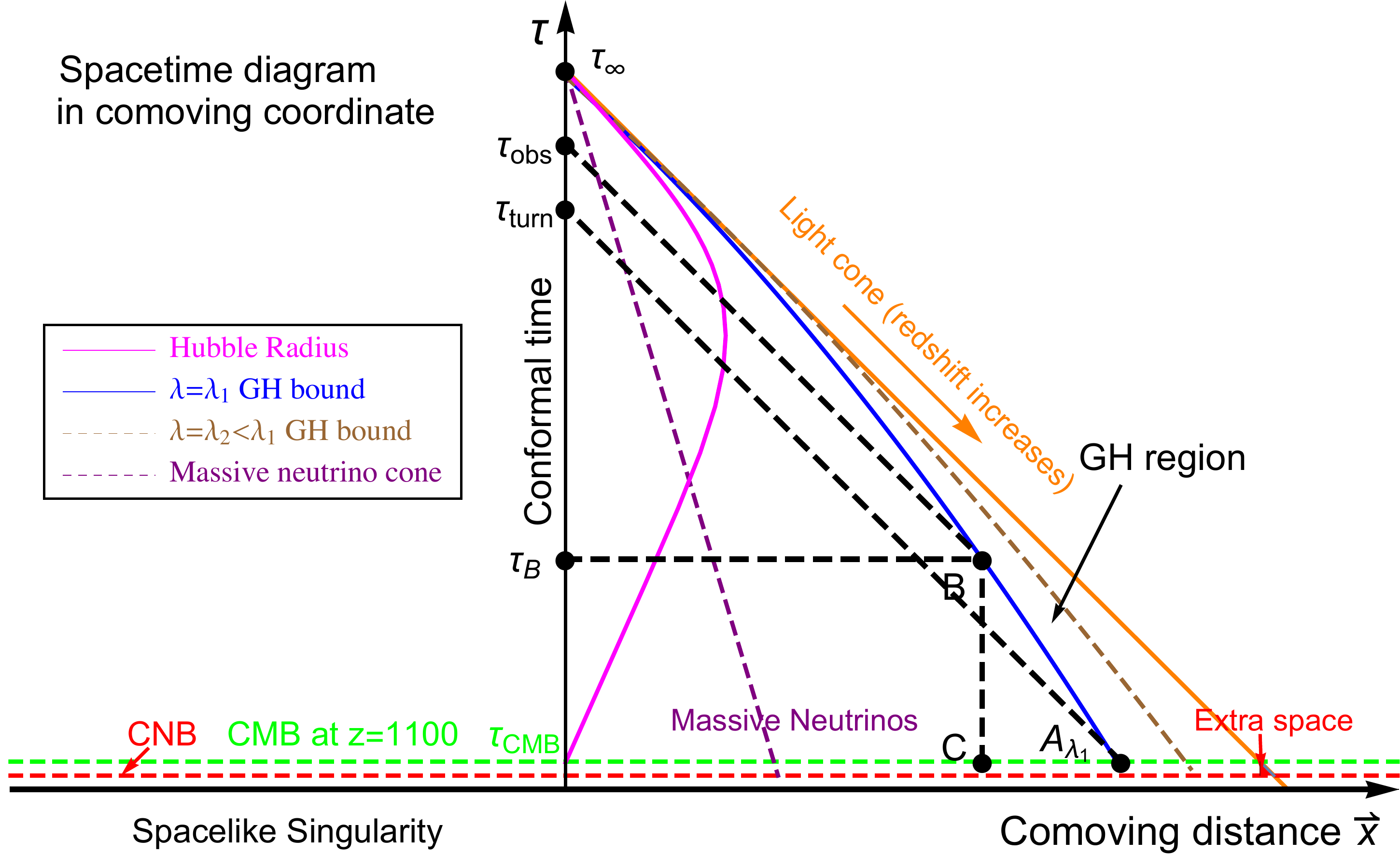}}
\caption{Spacetime diagram of the $\Lambda$CDM universe in comoving
coordinates, so that when $t\rightarrow \infty$, $\tau_{\infty}\simeq
4.4/H_{0}$ as a finite quantity. The horizontal black axis at the bottom is
the spacelike singularity at $t=0$, while the red dashed horizontal line is
the cosmic neutrino background (CNB) at $T \sim 1\,$MeV, and the green dashed
horizontal line is the CMB last-scattering
surface (LSS at $z=1100$).  The magenta line rising from $\protect\tau_{\rm
CMB}$ to $\protect\tau_{\infty}$ is the {\it comoving\/} Hubble radius.
The orange line with slope $-1$ is the past light cone
from the infinite future. The blue line is the
``Gibbons-Hawking'' (GH) bound for radiation with wavelength
$\lambda_{1}$ (and the GH bound for a shorter wavelength
$\protect\lambda_{2}$ is the brown dashed line to the right).
Light emitted from the CMB LSS (at point C) has its wavelength
stretched along its worldline until it reaches the GH bound (at point B)
and becomes ``screened.''  As the Universe
evolves, an observer can see a larger and larger space; however, once the
observation time reaches the turning point $\tau_{\rm turn}$ [light emitted at
$\lambda(A_{\lambda_{1}})\sim 1/H$], the observable length
starts to decrease.  Another phenomenon is that the last-scattering surface
of the cosmic neutrino background (CNB)
for sufficiently massive neutrinos is actually closer to us (purple
dashed line) \cite{Dodelson09} than the CMB.  However, for light neutrinos
we have access to some extra volume (red
region in the lower-right corner) of the universe along the past light cone
of the neutrinos.} \label{fig:spacetime}
\end{figure*}

\begin{figure*}[tbp]
\centerline{
\includegraphics[width=3.4in]{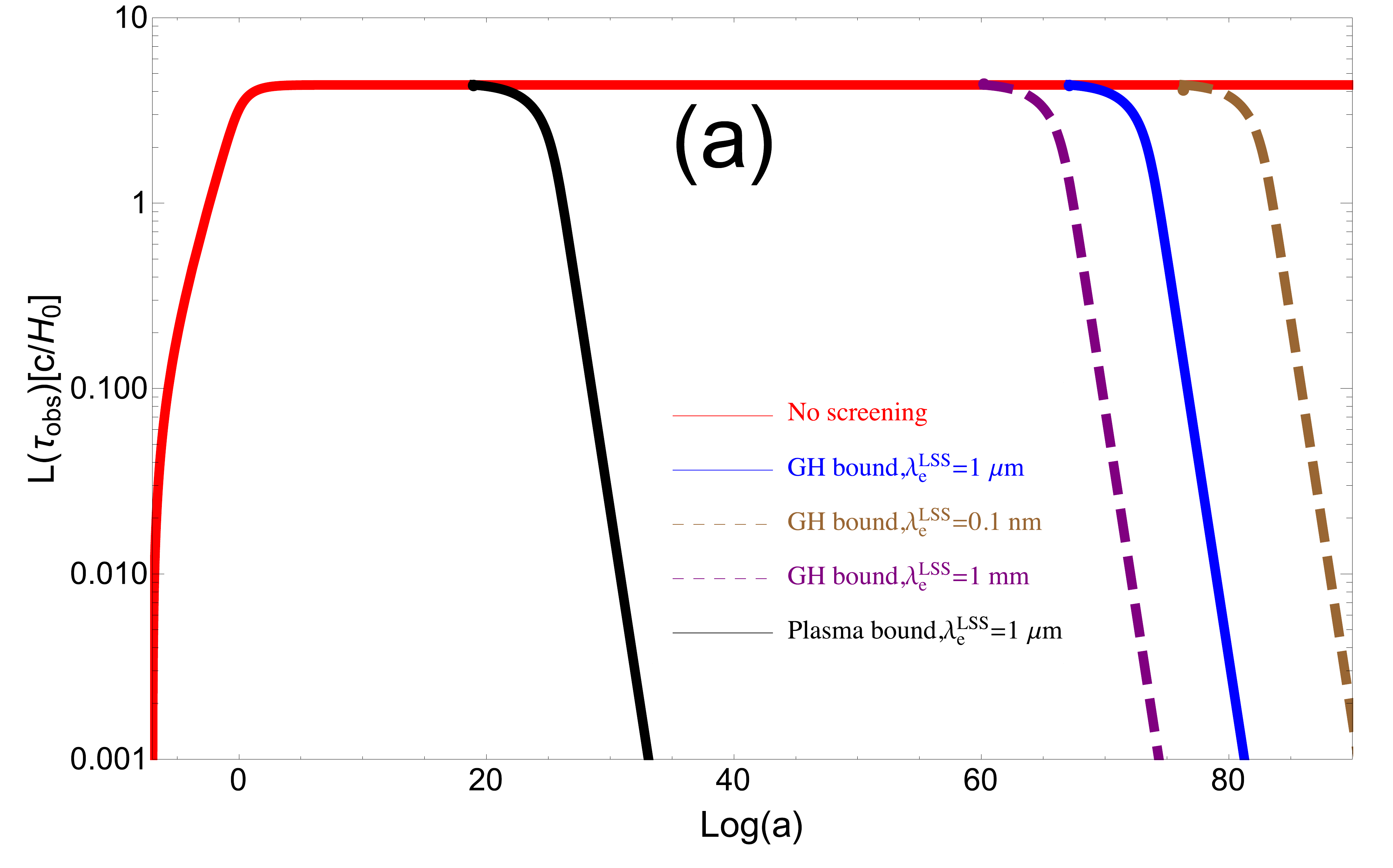}
\includegraphics[bb=0 20 600 300, width=2.9in]{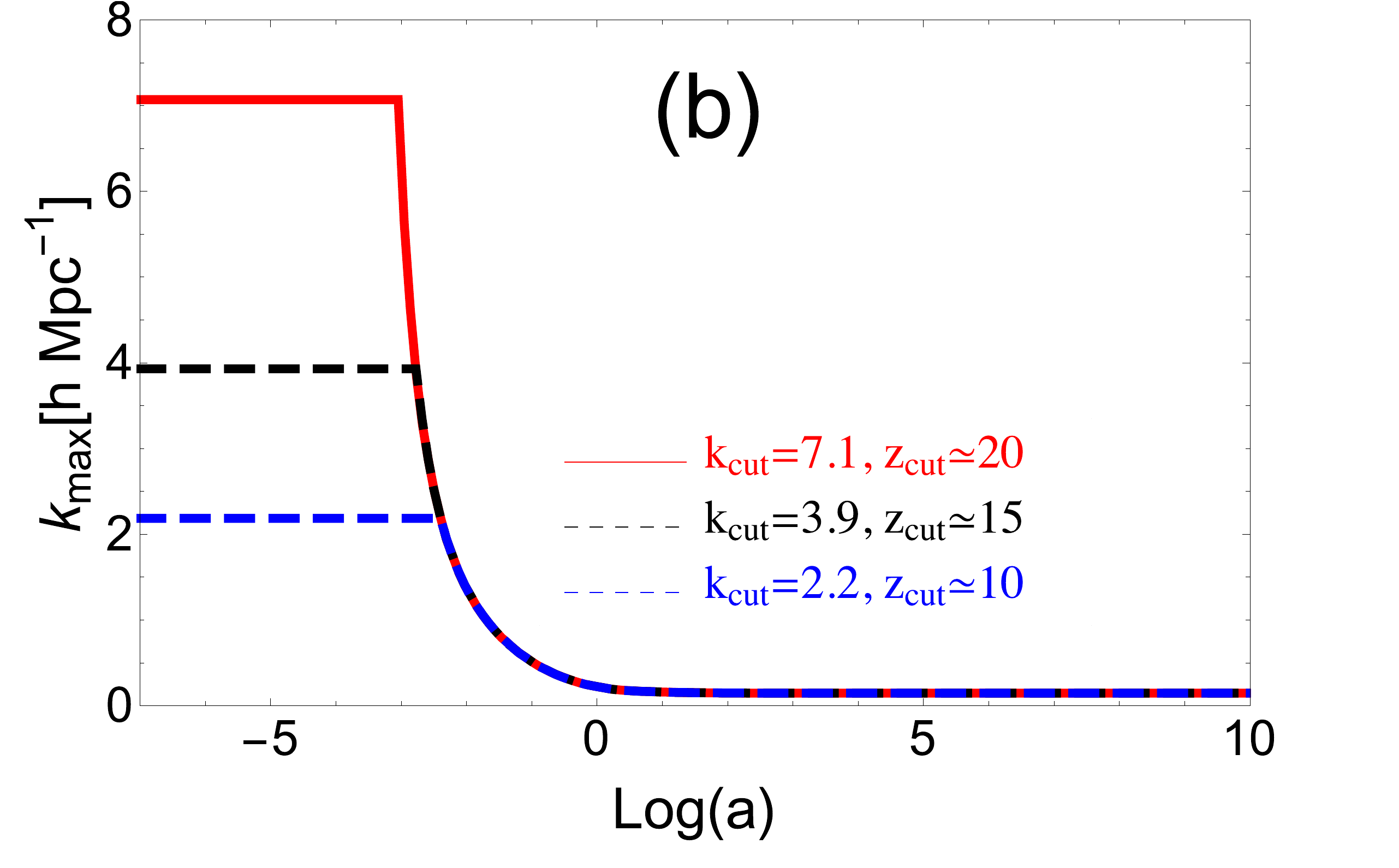}}
\caption{(a) Observable length scale (comoving frame) as
a function of logarithmic scale factor. (b) Effect of
different cutoffs in $k$ at early times due to various
possibilities for unknown small-scale physics, picking three example
values.} \label{fig:Lk}
\end{figure*}

Figure~\ref{fig:spacetime} shows a spacetime diagram from $t=0$
(neglecting an inflationary phase) to the infinite
future.  Due to the acceleration caused by dark energy, our comoving
Hubble length increases to its maximum at $z \simeq 10$, but afterwards
decreases as the dark energy starts to dominate. However, the
maximal volume of observable space is {\it not\/} limited to the
Hubble patch, since we are observing along the past light cone,
and so we will have access to larger and larger regions as time
passes.  It was suggested in Ref.~\cite{Loeb12} that we live at a time near the
maximum in the available number of modes, but in fact our observable patch
has {\it not\/} decreased from $z \simeq 10$ until now, and will not diminish
further in the future.  The comoving size of the observable Universe
at any time is an integral from the last-scattering surface
($\tau_{\ast}$) to the observed time ($\tau_{\rm obs}$):
\begin{eqnarray}
L(\tau_{\rm obs})=\int_{\tau_{\ast}}^{\tau_{\rm obs}} c \der \tau,
\label{eq:Lobs1}
\end{eqnarray}
which monotonically increases with time [shown as the red line in
Fig.~\ref{fig:Lk}(a)].  So even though the Hubble scale shrinks because
of the accelerating expansion, in the comoving frame we will 
be able to see more and more volume; however, the volume increases only
slowly in the far future, corresponding to a flattening of the red line
in Fig.~\ref{fig:Lk}(a) for $\log(a)>{\rm few}$.

\begin{figure*}[tbp]
\centerline{
\includegraphics[bb=0 0 756 486, width=3.4in]{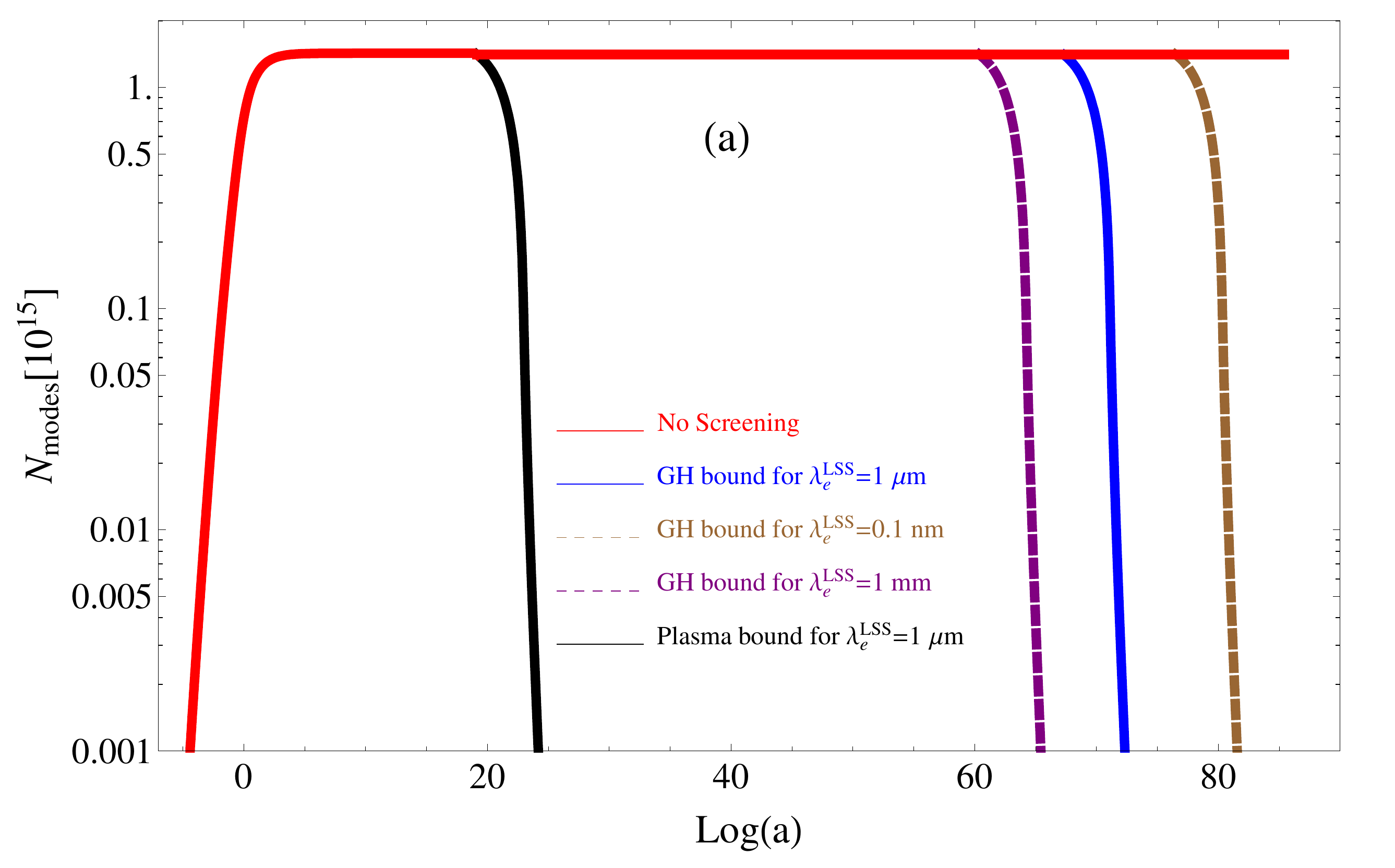}
\includegraphics[bb=0 0 780 500, width=3.4in]{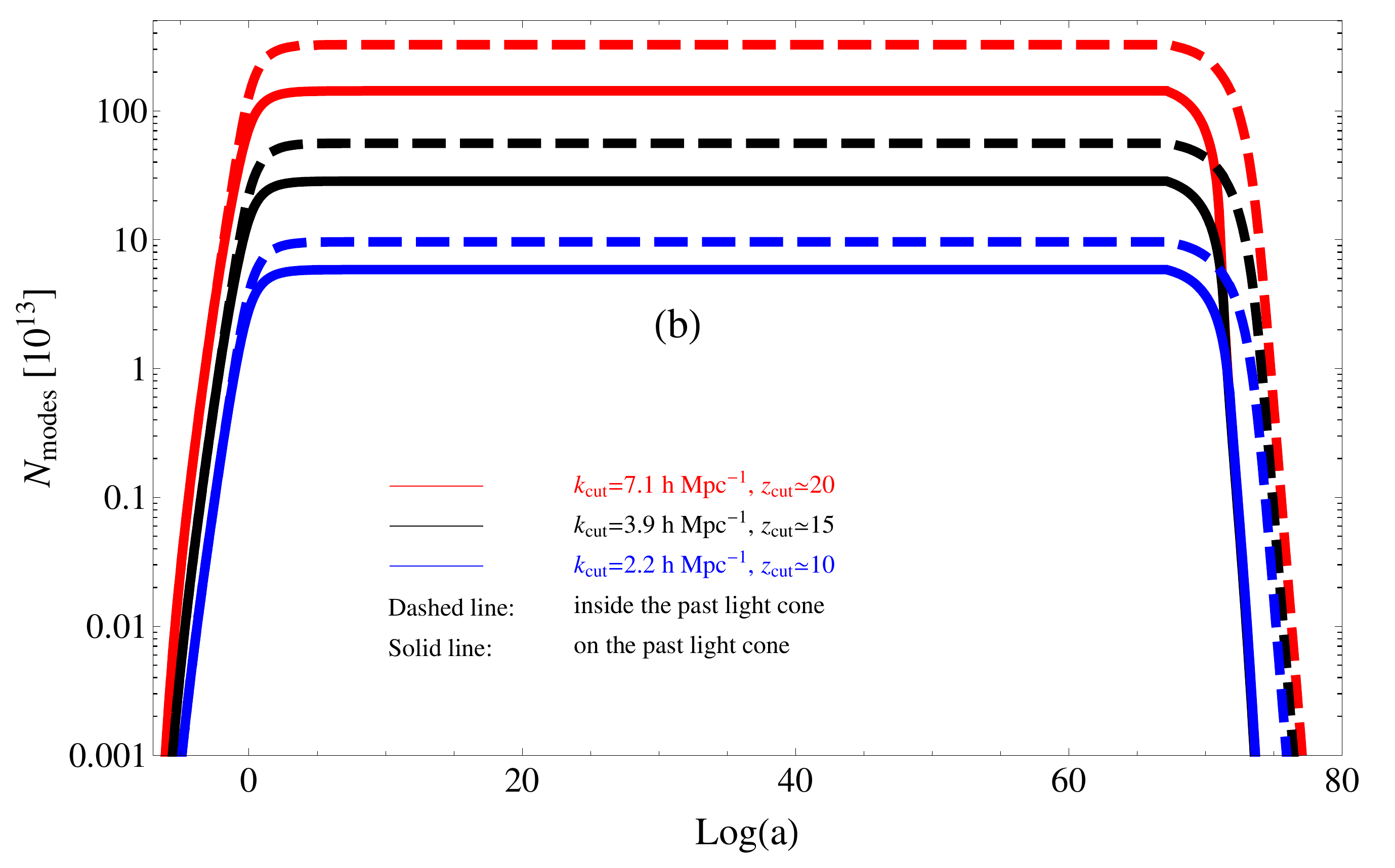}}
\caption{(a) Number of modes confined on the past light
cone that can be measured (here taking the early cutoff scale to
be $k_{\rm max}=7.1 \mpch$). (b) Sensitivity of
$N_{\mathrm{modes}}$ as a function of different cutoff scales.}
\label{fig:modes}
\end{figure*}

The CMB and other signals will certainly get {\it harder\/} to measure as they
redshift away into the $\Lambda$-dominated (de Sitter) phase, but it is
important to realize that they will not disappear entirely.  Nevertheless,
it is worth considering what happens in the far future, since eventually a
fundamental limit exists below which the CMB
temperature cannot be sensibly defined.  In an empty de
Sitter universe one would see a thermal bath with temperature (in natural
units) $T_{\rm dS}=H/2\pi=\sqrt{\Lambda/3}/2\pi$~\cite{Gibbons77}, and thus
after the CMB redshifts to $T_{\rm dS}$, it
becomes completely lost in the thermal noise of the de
Sitter background.  This is known as the ``Gibbons-Hawking
bound'' (GH) bound~\cite{Zibin07,Krauss00} and gives a limit in the very far
future of a purely $\Lambda$ model; it is equivalent to considering the time
when the typical wavelength of a CMB photon becomes equal to the Hubble scale,
so that one cannot sensibly define a temperature afterwards.

However, even well before this epoch, the CMB spectral peak will redshift
below the plasma frequency of the interstellar medium, and thus
would be screened from any observer within our Galaxy. The plasma
frequency in our Galaxy is around $1\,$kHz, which corresponds to a
wavelength of $\lambda_{\rm p}\simeq 3\times 10^{5}\,$m
\cite{Krauss07}.

These screening effects can limit the effectively observable size
of the Universe.  As an example, light with an emitted wavelength
($\lambda_{\rm e}$) originating from point C on the CMB LSS
in Fig.~\ref{fig:spacetime} can be observed at any time until
the light's wavelength is stretched
to the size of the Hubble length (point B), and then it becomes invisible
after $\tau_{\rm obs}$.  If light is emitted at point
A$_{\lambda_{1}}$ with wavelength comparable to
$cH^{-1}(a_\ast)$, then it is screened by the GH effect soon
after, and can never be detected. Therefore, as the Universe
expands, the observer can see a larger and larger volume of
space along the past light cone, until they reach the turning
point where $c/H=\lambda_{\rm e}(a_{\rm turn}/a_\ast)$. For the
plasma bound, this simply means replacing the Hubble length $c/H$
with the plasma wavelength $\lambda_{\rm p}$. After this time
$\tau_{\rm turn}$, the observable comoving length evolves as
\begin{equation}
L(\tau_{\rm obs})=\int^{\tau_{\rm obs}}_{\tau_{\rm e}} c \der
\tau, \label{eq:Lobs2}
\end{equation}
where $\tau_{\rm e}$ is the emission time corresponding to
scale factor $a_{\rm e}=(\lambda_{\rm e} a_{\rm obs})(c/H)^{-1}$.
In Fig.~\ref{fig:Lk}(a), we show the observable length scale as a
function of $\log(a)$, and we see that the observable length increases
until it is saturated by the effects of vacuum energy. In the far
future, the screening effect from the interstellar plasma or the
GH bound can decrease the observable size.
For example, light with peak CMB
wavelength at the LSS epoch corresponding to $\lambda \simeq
1\,\mu$m will have a turning point at $a\simeq e^{67}$, i.e.,
$1100\,$Gyr [blue line in Fig.~\ref{fig:Lk}(a)], while for a shorter wavelength
(e.g., the brown dashed line), the screening effect will happen later.

Of course, real observers billions of years into the future may need to
consider other events that are important for the evolution of their particular
location, such as the red giant phase of the Sun, or the merging
of the Milky Way and Andromeda galaxies.  This latter interaction would
significantly affect the ISM plasma and hence the observability of the CMB.
However, we neglect all such observer-specific effects here, and simply
consider the case in which the local conditions for the observer are continuously
stable over the full cosmic evolution, in the same spirit as previous
discussions of the future of cosmology (e.g., Ref.~\cite{Loeb12}).  In the
same vein, we are only considering the standard $\Lambda$CDM picture, with
the dark energy being exactly a cosmological constant, and with no additional
unconventional physics, such as future phase transitions.

\section{Total number of modes}
In the standard cosmological picture, nonlinear structure develops first
on small spatial scales (large $k$ values), where it erases any
memory of the initial conditions~\cite{Mobook}. One is free to pick any
reasonable definition for the scale of this nonlinearity; e.g.,
we can use $k_{\rm max}$ corresponding to the radius $R$ in a
spherical top-hat window
for which the root mean square of the filtered density field is unity; i.e.,
$\sigma^{2}(R)=
(1/2\pi^{2})\int P_{\rm m}(k)\left(3 j_{1}(kR)/(kR) \right)^{2}\der^{3} k=1$,
where $P_{\rm m}(k)$ is matter power spectrum and $j_{1}$ is the first-order
spherical Bessel function.  In Fig.~\ref{fig:Lk}(b), we
plot $k_{\rm max}\equiv2\pi/R$ as a function of $\log(a)$.
The universe becomes more clustered as it evolves, so that
the amplitude of $P_{\rm m}(k)$ grows and the size of the region
$R$ with rms unity fluctuations becomes bigger, until the
growth of structure is frozen out by the vacuum energy.  So
$k_{\rm max}$ is a decreasing function of $\log(a)$.

At early times $k_{\rm max}$ was relatively large, but
complicated baryonic physics (and potentially other exotic effects) at early
times on small scales can also limit our ability to probe
perturbation up to very high-$k$ modes, giving an effective
cutoff in $k$ at higher redshifts.  These effects include the
baryon gas pressure \cite{Loeb12}, the relative velocity between
baryons and dark matter~\cite{Tseliakhovich10}, the x-ray and
photo-ionization heating of intergalactic baryons from first-light
objects~\cite{Loeb04}, any warm dark matter thermal velocity~\cite{Bode01}, and the potential existence of primordial black
holes~\cite{Profumo06}.
To investigate the effect of this uncertainty, we take three typical
values of $k_{\rm cut}$ at corresponding redshifts, as shown in
Fig.~\ref{fig:Lk}(b), these being related to the specific nature of the dark
matter and baryonic feedback processes.

We can now calculate the total number of linear modes as
\begin{eqnarray}
 N_{\rm modes} &=& \frac{1}{(2\pi)^{3}} \int_{0}^{V(\tau_{\rm obs})} \der V
 \int^{k_{\rm max}}_{k_{\rm min}} \der^{3}k    \nonumber \\
 &\simeq& \frac{1}{6\pi^2} \int_{0}^{V(\tau_{\rm obs})} k^{3}_{\rm max}
(\tau_{\rm e})\, \der V. \label{eq:Nmode1}
\end{eqnarray}
Here, since $k_{\rm min}L_{\rm obs}\sim 1$, we
replace the lower limit of the $k$ integral with zero, and write
$k_{\rm max}$ as a function of the light emission time. We can
consider two distinct cases for observations carried out to recover the
power spectrum of fluctuations.

First we can assume that observations are confined to the
surface of our past light cone.  This would be valid for
traditional galaxy surveys~\cite{Eisenstein05}, gravitational lensing,
and 21-cm surveys~\cite{Lewis07}, for example. Thus for each observational time
$\tau_{\rm obs}$, when we integrate out to the maximum
observable length scale [Eqs.~(\ref{eq:Lobs1}) and
(\ref{eq:Lobs2})], we are integrating over all emission times out
to the LSS ($\tau_{\ast}$).  Figure~\ref{fig:modes}(a) shows the
evolution of the observable number of modes for such
observations, explicitly assuming a cutoff scale
$k_{\rm cut}\simeq 7.1 \mpch$ at $z\simeq 20$.  We find that, although it
will certainly become harder to extract cosmological information in
the future (because of the accelerated expansion), the number of
measurable modes still increases monotonically with time.  This
will be the case until screening effects (either ISM plasma or GH)
eventually prohibit observations in the very distant future.

As a second case, we can also consider observations that
probe the {\it inside\/} of our past light
cone.  Potential observations of this kind include the kinetic
Sunyaev-Zeldovich effect \cite{Sunyaev72}, spectral distortions
of the CMB due to scattered light~\cite{Kamionkowski97}, or anisotropies in
a massive neutrino background \cite{Hannestad10}.  In practice
little additional information is probably accessible
\cite{Bunn06}, but in principle, all of the interior of our
last-scattering surface could be observed in this way.  This represents
the fundamental limit of our ability to observe fluctuations due
to causality and the opacity prior to last scattering.  Since
such observations always measure the structure back to the LSS, we
can substitute $\tau_{\rm e}=\tau_{\ast}$ into
Eq.~(\ref{eq:Nmode1}) and take $k_{\rm max}(\tau_{\ast})$ out of
the integral, thus obtaining
$N_{\rm modes}=k^{3}_{\rm max}(\tau_{\ast})V_{\rm obs}/6\pi^{2}.$

Figure~\ref{fig:modes}(b) shows the number of modes for the two
cases (light cone surface and inside the light cone),
while varying the value of $k_{\rm cut}$.
One can see that the number of modes is
sensitive to the cutoff $k$ scale, and varies between about
$10^{13}$ and $10^{15}$.  Since (up to a small numerical
factor) the number of bits of information is essentially the same
as the number of accessible modes, the total amount of information
within the past light cone is somewhat bigger than the information
confined on the past light cone, but the two have a similar order
of magnitude.

\section{Information on cosmological parameters} 
We can also cast
the information question in terms of parameter uncertainties, i.e.,
we can ask how the available information maps into constraints in
determining the cosmological parameters. To obtain
an estimate of the total signal-to-noise ratio (SNR), we can simplify to a
situation where all parameters are determined except one, which we take to
be the overall amplitude of primordial fluctuations, $A_{\rm s}$. One can
then calculate the Fisher matrix (see e.g., Ref.~\cite{Tegmark97})
of the $A_{\rm s}$ parameter at any observational time.
Since $\Delta A_{\rm s}=F_{A_{\rm s}}^{-1/2}$, the total information
in $A_{\rm s}$ becomes \cite{Rimes05,Neyrinck06}
\begin{equation}
I_{A_{\rm s}} = \frac{1}{4\pi^{2}} \int^{V(\tau_{\rm obs})}_{0}\!
\der V \int_{0}^{k_{\rm max}(\tau_{\rm e})}\! k^{2} \der k
\left(\frac{\partial \ln P_{\rm m}(k)}{\partial \ln A_{\rm s}}\right)^{2}.
\label{eq:Fisher}
\end{equation}
Because $\partial \ln P_{\rm m}(k)/\partial \ln A_{\rm s}=1$, the above
equation immediately reduces to $I=N_{\rm modes}/2$; i.e., the
amount of information (in terms of ${\rm SNR}^2$) equals half
the number of Gaussian modes. Therefore, the ${\rm SNR}^2$ of
$A_{\rm s}$ measured at different cosmic epochs has a similar
shape to the evolution shown in Fig.~\ref{fig:modes}(a).  The
maximal SNR will depend on the specific value of the high-redshift cutoff
and details of the observational methods.  The precision in $A_{\rm s}$ as a
function of $k_{\rm cut}$ gives
$\delta A_{\rm s}/A_{\rm s}\simeq2.5\times10^{-7}k_{\rm cut}^{-1.5}$, with
$k_{\rm cut}$ in units of $h\,{\rm Mpc}^{-1}$.
The further one can probe into the small-scale regime at higher redshifts, the
tighter the parameter constraints that one can achieve.

For multiple parameters one can
consider that the SNR is effectively divided among them (with some being
better constrained than others of course).  It is possible that for a
particular power spectrum measurement, two parameters might be degenerate
with each other; in this case probing more fluctuation modes will improve
a combined parameter constraint, rather than the individual ones.

Note that the above calculation represents the ultimate precision
for parameter uncertainties, neglecting all of the
practical limitations of experimental noise and foreground
contamination.  For example, real 21-cm measurements will suffer
seriously from foreground emission (both Galactic and
terrestrial), which is generally $10^{5}$ (or more) times higher
than the underlying signal~\cite{Mao08}.  Nevertheless,
we can approach the ultimate SNR limit by being increasingly
innovative in future experimental designs, as well as by probing further
into the nonlinear regime.

\section{Other kinds of information} 
Astute readers may at this
point be wondering why we have assumed that the {\it only\/}
source of cosmological information is in the measurement of the
amplitude of modes in the fluctuating density fields.   Why are traditional
observations of background cosmological quantities (like $H_0$, luminosity
distance as a function of redshift, helium abundance, etc.) not included
in the discussion?  The answer is that although such observations are
extremely useful cosmological probes at the present day, as the data
improve dramatically, they will either provide relatively modest increments to
the amount of information, or will become equivalent to counting modes, like
the power spectrum measurements we have focused on.  Let us illustrate this
by considering a few specific examples.

\subsection{The Hubble constant and expansion rate}
The observation of type-Ia supernovae is a classical example of another kind
of cosmological data~\cite{Sullivan11}.
However, in terms of the ideal amounts of
information we have been discussing, these kinds of observation
offer only minimal additional constraints, as can be seen
directly from the conformal diagram in Fig.~\ref{fig:spacetime}.
At $\tau_{\rm obs}$, the supernova observations only provide information
on $H(z)$ along a relatively short radius (since it is hard to observe very
high-$z$ supernovae).  Additionally, the cosmological information is limited
by dispersion in the properties of standard candles, and so in practice
measurements of $H(z)$ will not decrease like $\sqrt{N}$ as we add supernovae.

But more importantly, as the data become of high enough signal-to-noise
ratio, one will have to appreciate that the {\it local\/} value of $H_0$ can
be slightly different from the global one, because of density variations.
Right now we have estimates of $H_0$ at the ${\rm SNR}\,{\simeq}\,50$ level,
and it is well known that ``cosmic variance'' on this number is a non-negligible
fraction of the current uncertainty \cite{Shi98,Wang98,Wojtak14,BenDayan}.
Hence, even an infinite number of local distance estimates will
give a result similar to only measuring ${\sim}\,10^4$ modes, in terms of
constraints on cosmological parameters.

In order to expand the amount of information, one has to go to higher
redshifts.  But then there are additional issues to consider, such as the
cosmic peculiar velocity field~\cite{Turnbull12} and the effects of
gravitational lensing~\cite{Amendola15}.
Precision measurements of luminosity distance will thus determine different
expansion histories in different directions.  This will degrade our ability
to improve the SNR of such measurements, although with sufficiently
ambitious surveys the correlated
fluctuations can be considered as a cosmological signal \cite{Hui06}.  Hence
endeavors to constrain the cosmological background 
parameters then become effectively measurements of cosmological power spectra,
where the information is determined by mode counting.

\subsection{Other direct constraints on background cosmology}
Another example of an astrophysical measurement of a ``background'' parameter
is the determination of helium or deuterium abundance at $z=0$.
One might natively imagine that it can be measured with infinite accuracy,
and hence provide a high-quality constraint on the baryon-to-photon ratio
$\eta_{\rm b}$, or cosmic baryon density $\Omega_{\rm b}h^{2}$,
independent of any modes.  However,
once one recalls that the Universe at the time of big bang nucleosynthesis (BBN) contains
perturbations in density (and hence $\Omega_{\rm b}h^2$, for
example) and that one measures the abundance in different regions of
today's Universe (actually along the light cone) where conditions will vary
from place to place (e.g., Ref.~\cite{Cook14}), then it becomes
clear that the precision of a single number is limited, and to make
further progress we are back to considering modes.  In addition, astration in
astrophysical systems (i.e., destruction in stars of light elements such as
deuterium), and experimental systematics may limit the precision of
determining primordial abundance~\cite{Steigman07,Cook14}.

In fact the same general argument applies to all such measurements, with the
exception of a direct determination of the cosmological constant, $\Lambda$.
If we could devise a laboratory experiment to determine the value of $\Lambda$,
then we could in principle make a measurement of this cosmological parameter
to arbitrary precision.  The reason such a measurement would be different is, of course, because the cosmological constant is a pure number, which does not
permit perturbations (and in that sense is more like a physical constant,
like the fine structure constant $\alpha$, than a cosmological parameter).
However, we know of no way to measure $\Lambda$ without
making measurements over cosmological scales.

\subsection{CMB distortions}
In the early Universe, energy stored in small-scale density perturbations is
dissipated through Silk damping, producing $\mu$- and $y$-type distortions
of the CMB spectrum~\cite{HuSS94,Chluba12}.
Observations with future experiments, such as {\it PIXIE\/}~\cite{Kogut11},
will enable us to measure such distortions and thereby place constraints on
the amplitude and shape of the primordial power spectrum at wave numbers
$k \lesssim 10^{4}\,{\rm Mpc}^{-1}$. The two types of distortion produced from
the small-scale power spectrum of primordial curvature perturbation
[${\cal P}_{\zeta}(k)\propto P(k)/k$] are~\cite{Chluba12}
\begin{eqnarray}
\mu & \simeq & {2.2} \int^{\infty}_{k_{\rm min}} {\cal P}_{\zeta}(k)
 \left[e^{-k/5400}-e^{-(k/31.6)^2}\right]^2 \der \ln k; \nonumber \\
y &\simeq & {0.4} \int^{\infty}_{k_{\rm min}} {\cal P}_{\zeta}(k)
 e^{-(k/31.6)^2}\der \ln k. \label{eq:mu-y}
\end{eqnarray}
One can see that measurements of the distortions provide integral
constraints on the primordial power spectrum in the regime
$1 \lesssim k \lesssim 10^{4}\,{\rm Mpc}^{-1}$, and hence a limit on
cosmological parameters that affect this power.  Although constraints on
nonstandard models are interesting, it is much harder to detect a signal from
the standard cosmology, where $\mu$ and $y$ distortions are expected to be only
a few $\times10^{-9}$.  Moreover, the integral constraint on small-scale
power seems unlikely to be competitive with direct measurements of power
spectra, and additionally there are other sources of spectral distortion
that will have to be dealt with first.  Nevertheless, this is a promising
way of probing a range of scales that is
inaccessible to current CMB anisotropy and large-scale structure observations.
Similar ``integral'' constraints on the small-scale power spectrum may also be
available from BBN~\cite{Jeong14}.

\subsection{Neutrinos and gravitational waves}
Despite the fact that electromagnetic experiments overwhelmingly dominate
present-day astronomy, one may also consider information
coming from other messengers, such as gravitational waves or
neutrinos.  However, this does not change our conclusions
fundamentally.  For neutrinos, unless they are genuinely massless
and travel at the speed of light, this allows us to see some extra
volume of space (see Fig.~\ref{fig:spacetime}).  The neutrino LSS is
actually closer to us than the CMB LSS (again see Fig.~\ref{fig:spacetime})
\cite{Dodelson09} and therefore we do {\it not\/} see more space using massive
neutrinos, but just fill in some of the observable volume.  Realistically
speaking of course, we have no expectation that measuring neutrino
anisotropies will ever be easy.

Gravitational wave detectors will have the potential to measure
absolutely calibrated distances to individual black hole binary sources,
which again is mainly measuring the expansion of the Universe~\cite{Hogan09}.
Perturbations in the primordial gravity waves, potentially detectable
through CMB observations of the so-called $B$-modes
(see, e.g., Ref.~\cite{Bicep2-Planck}), simply add a small fraction of the
amount of information coming from the other CMB power spectra
\cite{2dinfo}.

\section{Discussion}
Within the simplest cosmological picture the information
available to constrain the cosmology is dominated by the count of modes
in power spectra.  Approximately a $10^6$ CMB modes have already been
measured (see Ref.~\cite{PlanckI}) and the total number accessible is only a
factor of a few times larger than this.  To provide tighter constraints on
cosmological models one therefore has to go to three dimensions, where it
should be possible to measure more than $10^{12}$ modes.

Our current view of the large-scale Universe is that it is described by
just six parameters--but of course no one expects that things will
ultimately be that simple.  Certainly we expect neutrino density to be a
seventh parameter, and departures from the standard cosmology will also
presumably be found eventually (perhaps curvature, tensors, dark energy
equation of state, isocurvature perturbations, running spectral index, etc.),
including extensions that we
have not yet imagined.  Adding additional parameters to the background
cosmology would not change what we have described, but just add details to
the story.  It could also be that
the primordial perturbations break the Gaussian assumption that we have made--i.e., that we will eventually measure $f_{\rm NL}$--but if the
non-Gaussianity is weak, then it also does not alter the basic idea that
mode counting dominates our ability to define the cosmology that we live in.

However, there is a whole other side of cosmological constraints that we have
been neglecting, which concerns itself with structure formation; i.e., how
baryons populate dark matter halos and turn into galaxies.  When one uses
large three-dimensional surveys, one has to consider phenomenological descriptors
of nonlinear structure (bias parameters, halo model functions, etc.) as well
as the background parameters.  But in practice, this distinction may be rather
fuzzy.  Already, one of the standard cosmological parameters, namely the
reionization optical depth $\tau_{\rm ion}$, is unrelated to fundamental
processes, and in principle can be predicted from known physics.
The split is also complicated in practice because of the way that the
parameters are related to specific observables.  We expect this distinction to
become less clear in the future as more parameters are required to describe
the structure formation part of the cosmological picture.

These kinds of complications are highlighted in an interesting idea proposed
in Ref.~\cite{McDonald09}, using two distinct tracers of the same modes in
order to make apparently cosmic-variance-free determinations of cosmological
parameters.  This is certainly a promising technique that will be exploited to
obtain more information from large-scale surveys; however, it is ultimately
limited by the applicability of the purely linear bias approximation.  In
practice, it should enable us to (i) use redshift-space distortions to estimate
the velocity divergence power spectrum to its cosmic variance limit, rather
than having an error dominated by the uncertainty on the parameter $\beta$
(a combination of the growth rate and the bias of the tracer population);
and (ii) simultaneously give a mode-by-mode estimate of the ratio of the
biases of the two tracers and the parameter $\beta$, which are free of cosmic
variance.  The measurement of a velocity power spectrum is a pure background
cosmology determination, whose SNR comes from mode counting, and could one day
be very large, so that finding ways to reach the cosmic variance limit will
improve our ability to probe more cosmological modes.  On the other hand, it
seems doubtful that the SNR in the measurement of $\beta$
will ultimately be competitive, and moreover requires a full understanding of
the bias of galaxies, including an assumption that there is no stochastic
element.

Despite these reservations, we are sure that further methods will be developed
to use multiple tracers to directly probe more fluctuation modes,
while at the same time generating more complex models for structure
formation within the background and linear perturbations picture.  This is
the only way that we will fully exploit the information in the $>10^{12}$ modes
that should be accessible.

\section{Conclusions}
We have presented here a calculation of the total number of linear modes that
can be measured at different epochs of cosmic evolution. We separately consider
$N$ from the light confined on the past light cone and the light
scattering within past light cone. Our results show that the
total number of modes (which is equal to twice the amount of information on
parameter precision) increases monotonically; however, it will reach a
saturation point in the distant future due to $\Lambda$ domination, and
in the very far future will drop due to plasma or Gibbons-Hawking screening
effects.  The detailed precision obtainable on cosmological parameters depends
on the smallest scales that can be probed in the early Universe, with the
ultimate value for SNR being around $10^{7}$.
Cosmologists today appear to live at the epoch where
dark energy starts to dominate, so that measuring cosmological information
from structures might be easiest, and will become harder as we move billions of
years into the future.  Nevertheless, future cosmologists will always be able
to do better if they are inventive enough.

\textit{Acknowledgements}-- We thank Jim Zibin and Ali Narimani
for helpful discussions.  This research was supported by the
Natural Sciences and Engineering Research Council of Canada and by
the Canadian Space Agency.

\bibliographystyle{unsrt}
\bibliography{cosmo_3D_v7}

\begin{thebibliography}{10}

\bibitem{Hinshaw12}
G.~{Hinshaw}, D.~{Larson}, E.~{Komatsu}, D.~N. {Spergel}, C.~L. {Bennett},
  J.~{Dunkley}, M.~R. {Nolta}, M.~{Halpern}, R.~S. {Hill}, N.~{Odegard},
  L.~{Page}, K.~M. {Smith}, J.~L. {Weiland}, B.~{Gold}, N.~{Jarosik},
  A.~{Kogut}, M.~{Limon}, S.~S. {Meyer}, G.~S. {Tucker}, E.~{Wollack}, and
  E.~L. {Wright}.
\newblock {Nine-year Wilkinson Microwave Anisotropy Probe (WMAP) Observations:
  Cosmological Parameter Results}.
\newblock {\em \apjs}, 208:19, October 2013.

\bibitem{Planck2013I}
{Planck Collaboration, Planck 2013 results I}.
\newblock {Overview of products and scientific results}.
\newblock {\em \aap}, 571:A1, November 2014.

\bibitem{PlanckI}
{Planck Collaboration, Planck 2015 results I}.
\newblock {Overview of products and scientific results}.
\newblock {\em ArXiv e-prints}, February 2015.

\bibitem{Planck16}
{Planck Collaboration, Planck 2015 results XIII}.
\newblock {Cosmological parameters}.
\newblock {\em ArXiv e-prints}, February 2015.

\bibitem{2dinfo}
D.~{Scott}, D.~{Contreras}, A.~{Narimani}, and Y.-Z. {Ma}.
\newblock {The information content of cosmic microwave background
  anisotropies}.
\newblock {\em ArXiv e-prints}, March 2016.

\bibitem{Dodelson09}
S.~{Dodelson} and M.~{Vesterinen}.
\newblock {Cosmic Neutrino Last Scattering Surface}.
\newblock {\em Physical Review Letters}, 103(17):171301, October 2009.

\bibitem{Loeb12}
A.~{Loeb}.
\newblock {The optimal cosmic epoch for precision cosmology}.
\newblock {\em \jcap}, 5:028, May 2012.

\bibitem{Gibbons77}
G.~W. {Gibbons} and S.~W. {Hawking}.
\newblock {Cosmological event horizons, thermodynamics, and particle creation}.
\newblock {\em \prd}, 15:2738--2751, May 1977.

\bibitem{Zibin07}
J.~P. {Zibin}, A.~{Moss}, and D.~{Scott}.
\newblock {Evolution of the cosmic microwave background}.
\newblock {\em \prd}, 76(12):123010, December 2007.

\bibitem{Krauss00}
L.~M. {Krauss} and G.~D. {Starkman}.
\newblock {Life, the Universe, and Nothing: Life and Death in an Ever-expanding
  Universe}.
\newblock {\em \apj}, 531:22--30, March 2000.

\bibitem{Krauss07}
L.~M. {Krauss} and R.~J. {Scherrer}.
\newblock {The return of a static universe and the end of cosmology}.
\newblock {\em General Relativity and Gravitation}, 39:1545--1550, October
  2007.

\bibitem{Mobook}
H.~{Mo}, F.~C. {van den Bosch}, and S.~{White}.
\newblock {\em {Galaxy Formation and Evolution}}.
\newblock Cambridge University Press, May 2010.

\bibitem{Tseliakhovich10}
D.~{Tseliakhovich} and C.~{Hirata}.
\newblock {Relative velocity of dark matter and baryonic fluids and the
  formation of the first structures}.
\newblock {\em \prd}, 82(8):083520, October 2010.

\bibitem{Loeb04}
A.~{Loeb} and M.~{Zaldarriaga}.
\newblock {Measuring the Small-Scale Power Spectrum of Cosmic Density
  Fluctuations through 21cm Tomography Prior to the Epoch of Structure
  Formation}.
\newblock {\em Physical Review Letters}, 92(21):211301, May 2004.

\bibitem{Bode01}
P.~{Bode}, J.~P. {Ostriker}, and N.~{Turok}.
\newblock {Halo Formation in Warm Dark Matter Models}.
\newblock {\em \apj}, 556:93--107, July 2001.

\bibitem{Profumo06}
S.~{Profumo}, K.~{Sigurdson}, and M.~{Kamionkowski}.
\newblock {What Mass Are the Smallest Protohalos?}
\newblock {\em Physical Review Letters}, 97(3):031301, July 2006.

\bibitem{Eisenstein05}
D.~J. {Eisenstein}, I.~{Zehavi}, D.~W. {Hogg}, R.~{Scoccimarro}, M.~R.
  {Blanton}, R.~C. {Nichol}, R.~{Scranton}, H.-J. {Seo}, M.~{Tegmark},
  Z.~{Zheng}, S.~F. {Anderson}, J.~{Annis}, N.~{Bahcall}, J.~{Brinkmann},
  S.~{Burles}, F.~J. {Castander}, A.~{Connolly}, I.~{Csabai}, M.~{Doi},
  M.~{Fukugita}, J.~A. {Frieman}, K.~{Glazebrook}, J.~E. {Gunn}, J.~S.
  {Hendry}, G.~{Hennessy}, Z.~{Ivezi{\'c}}, S.~{Kent}, G.~R. {Knapp}, H.~{Lin},
  Y.-S. {Loh}, R.~H. {Lupton}, B.~{Margon}, T.~A. {McKay}, A.~{Meiksin}, J.~A.
  {Munn}, A.~{Pope}, M.~W. {Richmond}, D.~{Schlegel}, D.~P. {Schneider},
  K.~{Shimasaku}, C.~{Stoughton}, M.~A. {Strauss}, M.~{SubbaRao}, A.~S.
  {Szalay}, I.~{Szapudi}, D.~L. {Tucker}, B.~{Yanny}, and D.~G. {York}.
\newblock {Detection of the Baryon Acoustic Peak in the Large-Scale Correlation
  Function of SDSS Luminous Red Galaxies}.
\newblock {\em \apj}, 633:560--574, November 2005.

\bibitem{Lewis07}
A.~{Lewis} and A.~{Challinor}.
\newblock {21cm angular-power spectrum from the dark ages}.
\newblock {\em \prd}, 76(8):083005, October 2007.

\bibitem{Sunyaev72}
R.~A. {Sunyaev} and Y.~B. {Zeldovich}.
\newblock {The Observations of Relic Radiation as a Test of the Nature of X-Ray
  Radiation from the Clusters of Galaxies}.
\newblock {\em Comments on Astrophysics and Space Physics}, 4:173, November
  1972.

\bibitem{Kamionkowski97}
M.~{Kamionkowski} and A.~{Loeb}.
\newblock {Getting around cosmic variance}.
\newblock {\em \prd}, 56:4511--4513, October 1997.

\bibitem{Hannestad10}
S.~{Hannestad} and J.~{Brandbyge}.
\newblock {The Cosmic Neutrino Background anisotropy --- linear theory}.
\newblock {\em \jcap}, 3:020, March 2010.

\bibitem{Bunn06}
E.~F. {Bunn}.
\newblock {Probing the Universe on gigaparsec scales with remote cosmic
  microwave background quadrupole measurements}.
\newblock {\em \prd}, 73(12):123517, June 2006.

\bibitem{Tegmark97}
M.~{Tegmark}.
\newblock {Measuring Cosmological Parameters with Galaxy Surveys}.
\newblock {\em Physical Review Letters}, 79:3806--3809, November 1997.

\bibitem{Rimes05}
C.~D. {Rimes} and A.~J.~S. {Hamilton}.
\newblock {Information content of the non-linear matter power spectrum}.
\newblock {\em \mnras}, 360:L82--L86, June 2005.

\bibitem{Neyrinck06}
M.~C. {Neyrinck}, I.~{Szapudi}, and C.~D. {Rimes}.
\newblock {Information content in the halo-model dark-matter power spectrum}.
\newblock {\em \mnras}, 370:L66--L70, July 2006.

\bibitem{Mao08}
Y.~{Mao}, M.~{Tegmark}, M.~{McQuinn}, M.~{Zaldarriaga}, and O.~{Zahn}.
\newblock {How accurately can 21cm tomography constrain cosmology?}
\newblock {\em \prd}, 78(2):023529, July 2008.

\bibitem{Sullivan11}
M.~{Sullivan}, J.~{Guy}, A.~{Conley}, N.~{Regnault}, P.~{Astier}, C.~{Balland},
  S.~{Basa}, R.~G. {Carlberg}, D.~{Fouchez}, D.~{Hardin}, I.~M. {Hook}, D.~A.
  {Howell}, R.~{Pain}, N.~{Palanque-Delabrouille}, K.~M. {Perrett}, C.~J.
  {Pritchet}, J.~{Rich}, V.~{Ruhlmann-Kleider}, D.~{Balam}, S.~{Baumont}, R.~S.
  {Ellis}, S.~{Fabbro}, H.~K. {Fakhouri}, N.~{Fourmanoit},
  S.~{Gonz{\'a}lez-Gait{\'a}n}, M.~L. {Graham}, M.~J. {Hudson}, E.~{Hsiao},
  T.~{Kronborg}, C.~{Lidman}, A.~M. {Mourao}, J.~D. {Neill}, S.~{Perlmutter},
  P.~{Ripoche}, N.~{Suzuki}, and E.~S. {Walker}.
\newblock {SNLS3: Constraints on Dark Energy Combining the Supernova Legacy
  Survey Three-year Data with Other Probes}.
\newblock {\em \apj}, 737:102, August 2011.

\bibitem{Shi98}
X.~{Shi} and M.~S. {Turner}.
\newblock {Expectations for the Difference between Local and Global
  Measurements of the Hubble Constant}.
\newblock {\em \apj}, 493:519--522, January 1998.

\bibitem{Wang98}
Y.~{Wang}, D.~N. {Spergel}, and E.~L. {Turner}.
\newblock {Implications of Cosmic Microwave Background Anisotropies for
  Large-Scale Variations in Hubble's Constant}.
\newblock {\em \apj}, 498:1--10, May 1998.

\bibitem{Wojtak14}
R.~{Wojtak}, A.~{Knebe}, W.~A. {Watson}, I.~T. {Iliev}, S.~{He{\ss}},
  D.~{Rapetti}, G.~{Yepes}, and S.~{Gottl{\"o}ber}.
\newblock {Cosmic variance of the local Hubble flow in large-scale cosmological
  simulations}.
\newblock {\em \mnras}, 438:1805--1812, February 2014.

\bibitem{BenDayan}
I.~{Ben-Dayan}, R.~{Durrer}, G.~{Marozzi}, and D.~J. {Schwarz}.
\newblock {Value of H$_{0}$ in the Inhomogeneous Universe}.
\newblock {\em Physical Review Letters}, 112(22):221301, June 2014.

\bibitem{Turnbull12}
S.~J. {Turnbull}, M.~J. {Hudson}, H.~A. {Feldman}, M.~{Hicken}, R.~P.
  {Kirshner}, and R.~{Watkins}.
\newblock {Cosmic flows in the nearby universe from Type Ia supernovae}.
\newblock {\em \mnras}, 420:447--454, February 2012.

\bibitem{Amendola15}
L.~{Amendola}, T.~{Castro}, V.~{Marra}, and M.~{Quartin}.
\newblock {Constraining the growth of perturbations with lensing of
  supernovae}.
\newblock {\em \mnras}, 449:2845--2852, May 2015.

\bibitem{Hui06}
L.~{Hui} and P.~B. {Greene}.
\newblock {Correlated fluctuations in luminosity distance and the importance of
  peculiar motion in supernova surveys}.
\newblock {\em \prd}, 73(12):123526, June 2006.

\bibitem{Cook14}
R.~J. {Cooke}, M.~{Pettini}, R.~A. {Jorgenson}, M.~T. {Murphy}, and C.~C.
  {Steidel}.
\newblock {Precision Measures of the Primordial Abundance of Deuterium}.
\newblock {\em \apj}, 781:31, January 2014.

\bibitem{Steigman07}
G.~{Steigman}.
\newblock {Primordial Nucleosynthesis in the Precision Cosmology Era}.
\newblock {\em Annual Review of Nuclear and Particle Science}, 57:463--491,
  November 2007.

\bibitem{HuSS94}
W.~{Hu}, D.~{Scott}, and J.~{Silk}.
\newblock {Power spectrum constraints from spectral distortions in the cosmic
  microwave background}.
\newblock {\em \apjl}, 430:L5--L8, July 1994.

\bibitem{Chluba12}
J.~{Chluba}, A.~L. {Erickcek}, and I.~{Ben-Dayan}.
\newblock {Probing the Inflaton: Small-scale Power Spectrum Constraints from
  Measurements of the Cosmic Microwave Background Energy Spectrum}.
\newblock {\em \apj}, 758:76, October 2012.

\bibitem{Kogut11}
A.~{Kogut}, D.~J. {Fixsen}, D.~T. {Chuss}, J.~{Dotson}, E.~{Dwek},
  M.~{Halpern}, G.~F. {Hinshaw}, S.~M. {Meyer}, S.~H. {Moseley}, M.~D.
  {Seiffert}, D.~N. {Spergel}, and E.~J. {Wollack}.
\newblock {The Primordial Inflation Explorer (PIXIE): a nulling polarimeter for
  cosmic microwave background observations}.
\newblock {\em \jcap}, 7:025, July 2011.

\bibitem{Jeong14}
D.~{Jeong}, J.~{Pradler}, J.~{Chluba}, and M.~{Kamionkowski}.
\newblock {Silk Damping at a Redshift of a Billion: New Limit on Small-Scale
  Adiabatic Perturbations}.
\newblock {\em Physical Review Letters}, 113(6):061301, August 2014.

\bibitem{Hogan09}
C.~J. {Hogan}, B.~F. {Schutz}, C.~J. {Cutler}, S.~A. {Hughes}, and D.~E.
  {Holz}.
\newblock {Precision Cosmology with Gravitational Waves}.
\newblock In {\em astro2010: The Astronomy and Astrophysics Decadal Survey},
  volume 2010 of {\em Astronomy}, 2009.

\bibitem{Bicep2-Planck}
{BICEP2/Keck and Planck Collaborations}.
\newblock {Joint Analysis of BICEP2/Keck Array and Planck Data}.
\newblock {\em Physical Review Letters}, 114(10):101301, March 2015.

\bibitem{McDonald09}
P.~{McDonald} and U.~{Seljak}.
\newblock {How to evade the sample variance limit on measurements of
  redshift-space distortions}.
\newblock {\em \jcap}, 10:007, October 2009.

\end{thebibliography}

\end{document}